\documentclass[aip,apl,preprint]{revtex4-1}
\usepackage{graphicx}
\usepackage{color}
\usepackage{url}
\usepackage{amssymb}
\usepackage{float}
\usepackage{graphicx} 
\usepackage{amsmath}
\usepackage[margin=0.5in]{geometry}
\usepackage[dvipsnames]{xcolor}
\usepackage{soul}
\usepackage{multirow}
\usepackage{newtxtext}
\usepackage{booktabs}
\usepackage{nicefrac}

\usepackage[mathlines]{lineno}% Enable numbering of text and display math
\newcommand{\BostonCollege}{Department of Physics, Boston College, Chestnut Hill, MA, USA}

\begin{document}
\title{Transverse Photoresistivity from Photothermal Current Deflection in Metal Films}

\author{Piyush Sakrikar}
\author{Vincent M. Plisson}
\author{Cameron Grant}
\author{Dylan Rosenmerkel}
\author{Gabriel Natale}
\author{Michael Geiwitz}
\author{Ying Ran}
\author{Krzysztof Kempa}
\author{Kenneth S. Burch}
\email{burchke@bc.edu}
\affiliation{\BostonCollege}
\date{\today}

\begin{abstract}
Quantum geometry in centrosymmetric systems has motivated the search for photocurrent responses beyond second order. In particular, electric field-induced nonlinear responses may also enable intrinsic polarization-sensitive optical detectors. Despite numerous efforts, clear methods are still needed to remove experimental artifacts, separating intrinsic from extrinsic effects, and disentangling linear responses from their higher-order counterparts. Here, we provide a systematic study of fabrication and measurement techniques to remove external artifacts in photoelectronic responses. This reveals a previously hidden photothermoelectric response in the transverse photoresistivity of symmetric thin films of simple metals. We identify its origin in thermal gradients producing current deflection and determine the device design and measurement parameters to minimize extrinsic effects that arise in photoinduced electronic responses.
\end{abstract}
\maketitle

The potential for novel optoelectronic devices\cite{liu2020semimetals} and revealing quantum geometry\cite{ma2021topology} has driven increasing interest in nonlinear photocurrent generation. Specifically, the connection between the material's quantum metric and second- or higher-order electronic response when illuminated with light. While nonlinear photocurrent measurements can provide significant insight into the underlying physics, they are complex measurements and require careful analysis.
 
At second-order, the two primary intrinsic mechanisms for generating a photocurrent ($\mathrm{J_i}$) along a given crystal direction $i$ are the Bulk Photovoltaic effect (BPVE) and the circular photogalvanic effect (CPGE). These depend on the applied light field $\mathrm{\mathbf{E}}$ as $\mathrm{J_i\ =\ \sigma_{ijk}^{(2)} \mathbf{E}_j\mathbf{E}_k^*}$, illustrating key characteristics of the response. Specifically, (1) they scale linearly with the light intensity ($\mathrm{I=|E|^2}$), (2) they are observable only in materials with broken inversion symmetry, and (3) they depend on the polarization state of incident light. The intrinsic and ultrafast nature of the polarization dependence offers new opportunities for polarization-dependent photodetection throughout the electromagnetic spectrum. \cite{osterhoudt2019colossal,ma2021topology,bernevig2018recent,yan2017topological} In addition, these effects are closely related to topological quantities such as the Berry connection. \cite{morimoto2016topological,chan2017photocurrents}

However, other photoinduced mechanisms complicate measurements of BPVE and CPGE. As such, it is crucial to be aware of all possible artifacts that one might encounter in such experiments, especially responses near the edges of the device.\cite{wang2023visualization,karch2011terahertz} In addition, the need for non-centrosymmetric materials severely limits the use of CPGE and BPVE in applications, as well as our ability to probe quantum geometry in most materials. Thus, there is increasing interest in exploring current-induced nonlinear effects. For example, applying a weak perturbation (i.e., an electric field) that breaks the inversion and time-reversal symmetries should unlock an intrinsic and effective second-order response. Indeed, current and voltage bias allowed the observation of second harmonic generation (SHG) in NbN,\cite{nakamura2019infrared} electrically gated $\mathrm{(Bi_{1-x}Sb_{x})_{2}Te_3}$ films,\cite{pan2017helicity} and the photohall effect in graphene.\cite{an2013enhanced,bykov2012second,mciver2020light} However, as with second-order effects, current-induced photoelectric responses can suffer from extrinsic responses due to device and experimental design that allows unwanted thermal and electrical artifacts. 

\begin{figure}[t]
    \centering
    \includegraphics[width=.9\textwidth]{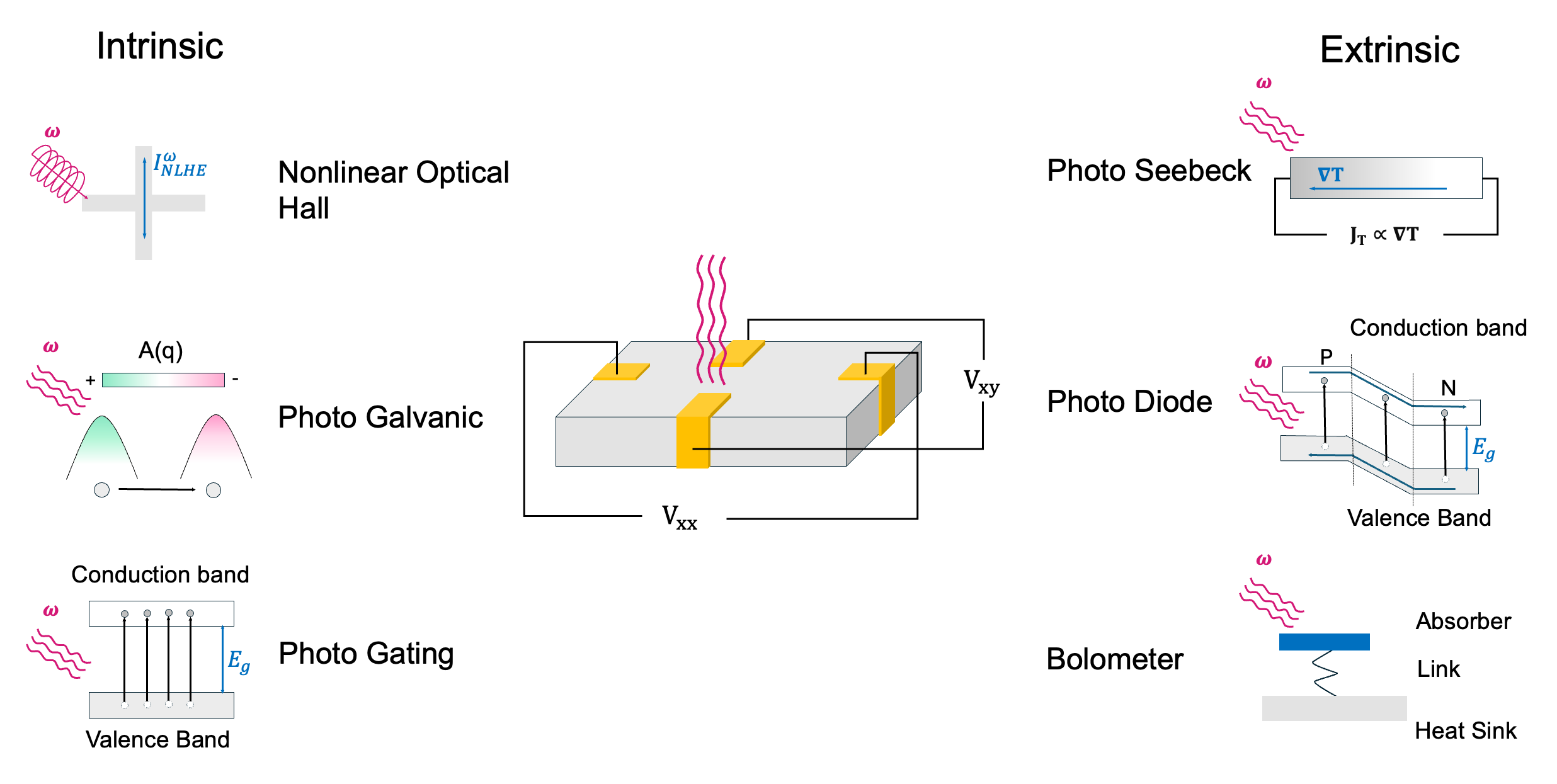}
    \caption{Schematic diagram of Optoelectronic responses: The optically induced change in resistance is measured via the longitudinal ($\mathrm{V_{xx}}$) and transverse ($\mathrm{V_{xy}}$) voltages. The left side of the figure shows a few common intrinsic nonlinear responses. The right side highlights extrinsic effects.}
    \label{fig: Responses to Light}
\end{figure}

Some of these commonly observed intrinsic and extrinsic contributions to photoinduced electric responses are summarized in FIG \ref{fig: Responses to Light}. Here, the schematic in the center represents a measurement consisting of a device fabricated with electrodes and light normally incident on the surface. The most commonly observed extrinsic response is due to the photo-Seebeck effect, which results from a thermal gradient induced by light absorption combined with an interface with changing Seebeck coefficients.\cite{xu2010photo} The photodiode effect occurs when photons generate electron-hole pairs that are separated by built-in electric fields due to doping inhomogeneity or Fermi-level pinning at the contacts. In the bolometric effect, light absorption leads to a localized increase in temperature, altering device resistance. Each of these effects depends linearly on the intensity of the applied light and is generally polarization-independent. Thus, they can mask the intrinsic contributions shown on the left of FIG \ref{fig: Responses to Light}. Namely, the nonlinear optical Hall effect,\cite{ma2019observation,sodemann2015quantum} photo-galvanic or BPVE, and photogating.\cite{furchi2014mechanisms} Furthermore, these can become nonlinear in intensity or have nontrivial polarization dependence depending on the material employed, device design, non-uniform illumination, and temperature.

Motivated by these concerns, we focus on device design, preparation, and measurement techniques that minimize extrinsic artifacts. Specifically, we explore the mid-infrared (MIR) spectral excitation in thin metal films. This range is of general interest for sensing, falls well above the free carrier response of most topological semimetals, and thus has been typically used to probe their quantum geometry.\cite{chan2017photocurrents,osterhoudt2019colossal} It is also a common wavelength for exciting phonons in substrates and is thus likely to induce unwanted thermal responses. With this in mind, we find that devices with fourfold rotational symmetry and contacts made from the same material as that under study exhibit dramatically reduced thermal artifacts. We also discuss the importance of carefully choosing a chopping mechanism to eliminate higher-order responses when measuring with a lock-in. Lastly, we show that these optimizations allow us to observe a transverse photoresistivity that results from current deflection due to nonuniform thermal illumination. 

%%%%%%%%%%%%%%%%%%%%%%% RESULTS SECTIONS %%%%%%%%%%%%%%%%%%%%%%%%%%%%%%%%%%%%%%

\begin{figure}[h]
    \centering
    \includegraphics[width=.9\textwidth]{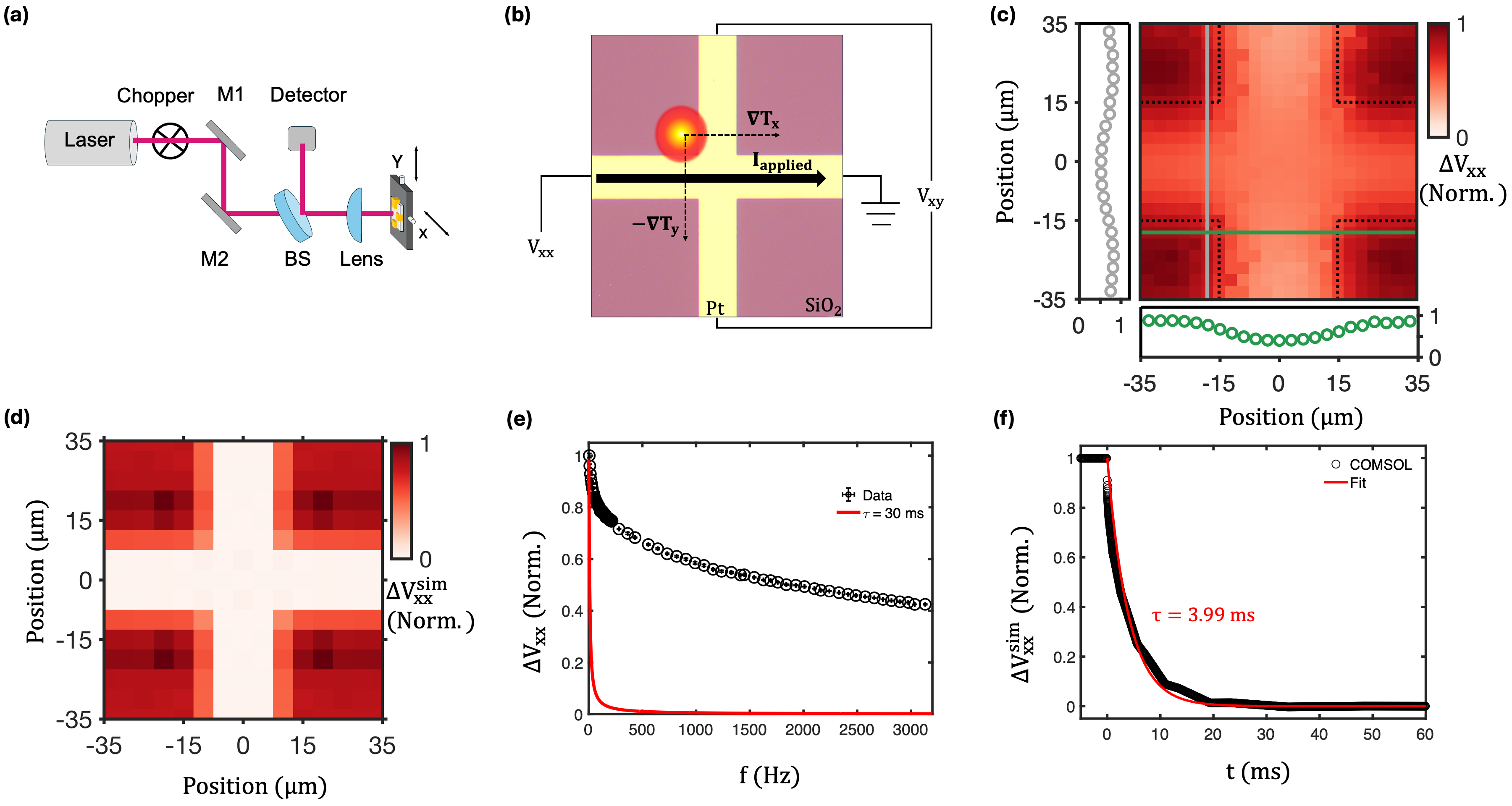}
    \caption{ (a)  MIR laser, 4.8 or 10.6 $\mathrm{\mu}$m, is focused on the sample while photovoltage and reflectance maps are simultaneously collected by moving the device through the fixed spot. (b) Microscope image of the metal cross-device overlaid with wiring schematic of the experiment to measure third-order longitudinal ($\mathrm{ V_{xx}}$) and perpendicular ($\mathrm{V_{xy}}$) Voltages. The bright red spot shows the laser illumination and resulting temperature gradient components ($\mathrm{\nabla T}$). (c) A map of the longitudinal photovoltage ($\mathrm{\Delta V_{xx}}$) normalized by its maximum value, with the sample denoted by black dotted lines. The solid lines, grey (vertical) and green (horizontal), represent the line cut profile of the photovoltage response shown outside the plot. The largest response is generated when the laser is on the SiO$_2$/Si substrate. (d) COMSOL simulation map of the normalized longitudinal photovoltage ($\mathrm{\Delta V^{sim}_{xx}}$) with equivalent results to the measurement. (e) Normalized $\mathrm{\Delta V_{xx}}$ versus chopper frequency showing $\tau \geq 1/(2\pi\times5~Hz)~ \approx ~30$ ms. (f) COMSOL simulation of $\mathrm{\Delta V^{sim}_{xx}}$ temporal decay with $\tau = 3.99$ ms.}
    \label{fig: Longitudinal}
\end{figure}

We employ a MIR quantum cascade laser, as shown by the schematic in FIG \ref{fig: Longitudinal}(a), operating in continuous wave mode with power stability of 1\%. An optical chopper modulates the light, and a lock-in amplifier detects various harmonics of the resulting voltage, with noise below 450 nV.  Our laser then passes through a beamsplitter, which lets us collect the reflected light. Finally, the light passes through a lens and onto our device. The device is mounted on a motorized X-Y stage for mapping without changing the angle of incidence. The latter could lead to additional nonlinear terms (photon drag) and/or thermal responses due to inhomogeneous excitation. \cite{wang2023visualization,ma2021topology,glazov2014high} 

Our electrical configuration is shown in FIG \ref{fig: Longitudinal}(b) and is motivated by electric field-induced second-order measurements in centrosymmetric materials. Here, we apply a constant current along the x-direction of the 45 nm thick simple metal film and measure the photoinduced longitudinal ($ V_{xx}$) and transverse  ($V_{xy}$) voltages. We employed platinum and gold, noting that they are both centrosymmetric, have no resonances in the MIR, are well studied, easily fabricated, and allow contacts made from the same material as the device being studied.

Our first objective is to eliminate any signal contributions from the photo-Seebeck effect. This is the most commonly observed extrinsic effect in photocurrent measurements and is particularly exacerbated when using IR wavelengths due to significant heating from the laser. Three main factors lead to a large Seebeck response: laser-induced inhomogeneous thermal gradients, anisotropy in the material's Seebeck coefficient, and a change in the Seebeck coefficient at the contacts.\cite{ma2021topology, wang2023visualization,glazov2014high} In most experiments, having an interface with a changing Seebeck coefficient is inevitable due to the different materials at the contact. This leaves us with only one choice: optimizing devices and measurement techniques to minimize the thermal gradient at the two contact junctions. Consequently, there are two ways to solve this problem. The first is to remove any thermal gradient generated by the laser altogether, which can be done by using a laser spot that is much larger than the device. Indeed, this technique has been successful, \cite{glazov2014high} but requires high-power light sources to reach the fluence levels needed for nonlinear responses. Alternatively, we have chosen to fabricate highly symmetric samples to allow for even heat loads and to extend the contact interfaces far enough from the center of the device so that they are outside the region of local heating from the laser. With no bias current, this approach was successful in eliminating the photo-Seebeck effect when the MIR beam was illuminated over the device. Having eliminated the first-order Seebeck effect through our design,  we now turn to effect of a DC bias current. This result, as well as all those reported in this manuscript were reproduced in at least two devices, each measured on five different days.  

We begin by checking the longitudinal photovoltage with a DC current applied along the $x$ direction while measuring the resulting AC voltage due to the chopped laser ($\mathrm{\Delta V_{xx}}$). The representative response of the platinum device to a 4.8 $\mu$m laser versus its position is shown in FIG \ref{fig: Longitudinal}(c). To focus on relative changes induced by the laser, without complications due to uncertain optical constants, angle of incidence, spot size, and thermal resistance, all presented maps are normalized to the maximum $\mathrm{\Delta V_{xx}}$. As seen in the line cuts in FIG \ref{fig: Longitudinal}(c), the photoinduced voltage does not change sign as the laser passes over the center of the device. However, such sign changes are typically observed from photo-Seebeck effects due to the change in the direction of the thermal gradient. Instead, we see a nearly constant positive $\mathrm{\Delta V_{xx}}$ along the x-direction that is weakened when the laser passes over the platinum. As we move the laser spot in the y-direction, the response decreases as the distance from the current-carrying wire increases. Consistent with a bolometric response, the photovoltage scales linearly with the applied current. We note that similar responses were seen in Au devices and illumination at 10.6 $\mathrm{\mu}$m wavelength.

As expected from a thermal effect, the lower $\mathrm{\Delta V_{xx}}$ when the laser directly illuminates the metal is easily explained by the platinum's high IR reflectivity.\cite{polyanskiy2024refractiveindex} However, since the substrate consists of $285$ nm SiO$\mathrm{_2}$ on $p^+$Si, the laser heating is stronger due to enhanced light absorption.\cite{kitamura2007optical} Additionally, SiO$_2$ has approximately two orders of magnitude lower thermal conductivity (1.1 W m$^{-1}$ K$^{-1}$)\cite{he2025thermal} compared to $p^+$Si (110 W m$^{-1}$ K$^{-1}$)\cite{fan2022systematic} and Pt (72 W m$^{-1}$ K$^{-1}$),\cite{ho1972thermal} causing the heat to be trapped and diffuse slowly to the surrounding area. This results in a larger temperature increase in the adjacent platinum compared to when the laser is directly on the wire. Consequently, as the beam moves away from the current-carrying wire, the response decreases due to diminished heating across the wire. To confirm if the origin of this response is indeed dominated by platinum heating from optical absorption in the SiO$_2$/Si substrate, we performed COMSOL simulations to replicate our measurement. We employ a Gaussian heat source to mimic the laser-induced heating and solve the steady-state heat diffusion equation in our sample geometry. The details of the simulation, including the mesh size and tolerance, are in the supplementary material, and the results are shown in FIG \ref{fig: Longitudinal}(d). We find good agreement between the simulated response and our experimental measurements.

If this is the case, then the $\mathrm{\Delta V_{xx}}$ signal should be sensitive to the thermal relaxation time. We therefore modulated the chopper frequency and recorded the signal as shown in FIG \ref{fig: Longitudinal}(e). The response decreases with increasing chopper frequency, and we expect the typical bolometric frequency relation $\mathrm{(1 + (2\pi f\tau)^2)^{-1/2}}$,  where f is the frequency and $\mathrm{\tau}$  the relaxation time.\cite{richards1994bolometers} The mechanical chopper only permits us to reach modulation frequencies from 5 Hz to 3 kHz, which was not slow enough to observe the maximum response or the roll-off behavior. Hence, the fitting did not perfectly describe the data. The experiment yields a much longer effective timescale $\mathrm{\tau \geq 1/(2\pi\times5~Hz)~ \approx ~30}$ ms, indicated by the solid line, than our chopping range. To validate our thermal interpretation, we performed time-domain COMSOL simulations, shown in FIG \ref{fig: Longitudinal}(f), to model the temporal evolution of the steady-state photovoltage response after the heat source is switched off. The simulated $\mathrm{\Delta V^{sim}_{xx}}$ response exhibits an exponential decay with time constant $\mathrm{\tau} = 3.99$ ms. The discrepancy between experimental and simulated time constants likely arises from the simplified thermal boundary conditions in our model and the complex heat dissipation pathways in the actual device geometry. Nonetheless, the strong dependence of the longitudinal data on the modulation frequency is consistent with it being bolometric.

\begin{figure}[h] 
    \centering
    \includegraphics[width=0.9\textwidth]{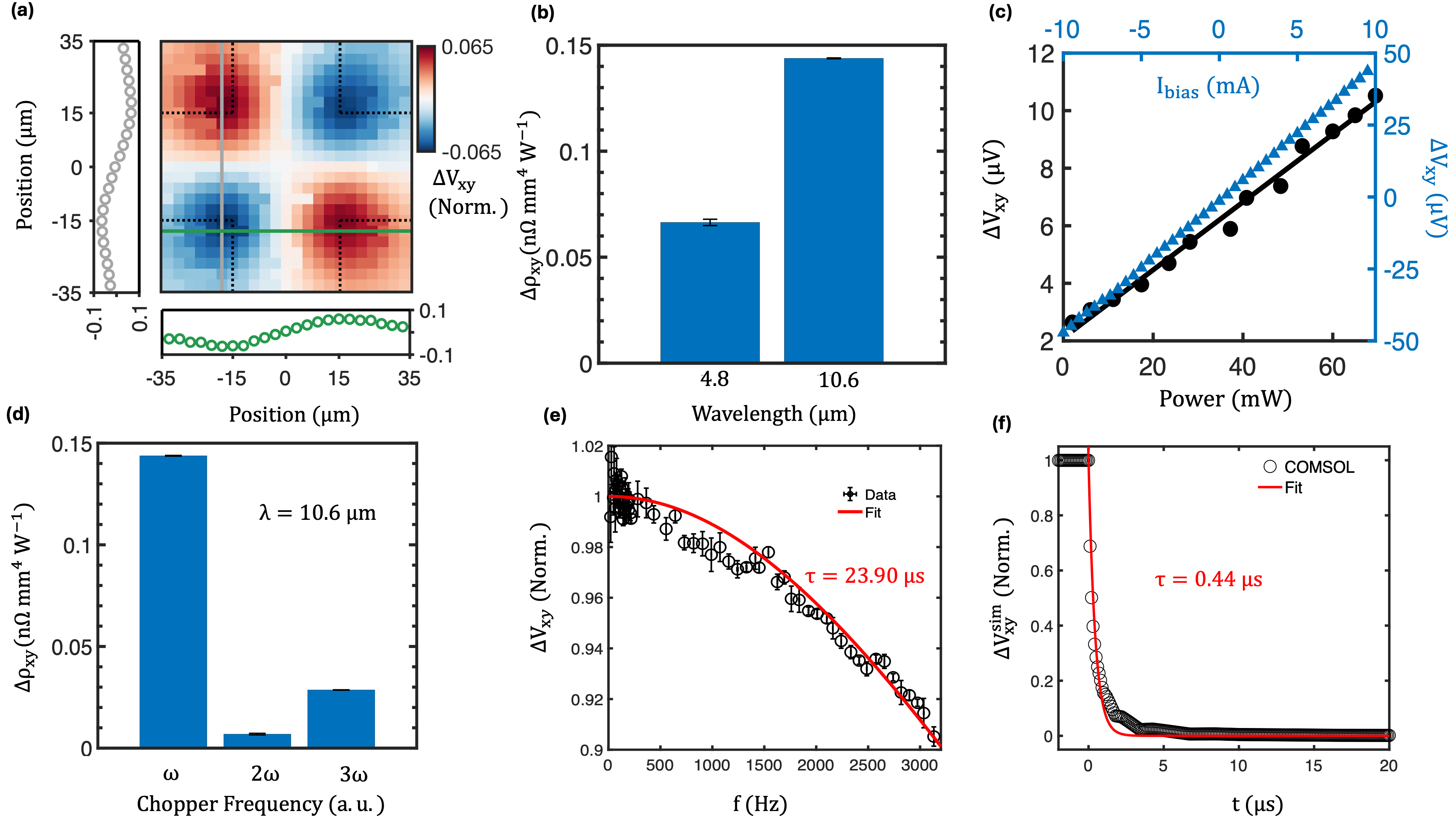}
    \caption{(a) A map of the transverse photovoltage normalized by the maximum longitudinal response, with the device denoted by black dotted lines. The solid lines, grey (vertical) and green (horizontal), represent the line cut profile of the photoresponse shown outside the plot. (b) Histogram plots of the peak transverse response of the 30$\mathrm{\mu}$m platinum device as a function of incident wavelength. Error bars represent the standard deviation of peak values measured across current hotspots. (c) The transverse response is linear in both applied current and laser Power. The solid line is a linear fit to the latter. (d)  Histogram plots of the peak transverse response of the 30 $\mathrm{\mu}$m platinum device as a function of the chopper modulation frequency. Error bars represent the standard deviation of peak values measured across current hotspots. Frequency-dependent thermal response and relaxation dynamics. (e) Normalized transverse photovoltage $\mathrm{\Delta V_{xy}}$ versus chopper frequency with fit to $\mathrm{(1 + (2\pi f\tau)^2)^{-1/2}}$ yielding $\mathrm{\tau = 23.90\ \mu}$s. (f) COMSOL time-domain simulation of $\mathrm{\Delta V_{xy}^{sim}}$ decay, with exponential fit giving $\mathrm{\tau = 0.44}$ $\mathrm{\mu}$s. }
    \label{fig: Transverse}
\end{figure}

Next, we turn to the transverse photovoltage ($\mathrm{\Delta V_{xy}}$) experiments that could directly connect to quantum geometry. In such experiments, one does not anticipate a transverse response to heat in time-reversal and inversion symmetric materials. FIG \ref{fig: Transverse}(a) shows the transverse photovoltage normalized to the maximum response observed $\mathrm{\Delta V_{xx}}$ from the same device shown in FIG \ref{fig: Longitudinal}(b). Despite the platinum films being multi-domain and employing linearly polarized light, we observe a previously unreported $\mathrm{\Delta V_{xy}}$. In this transverse measurement geometry, there is no signal when the laser is positioned exactly in the center of the cross. Similar to the longitudinal response, the largest signals are observed when the laser illuminates the SiO$_2$/Si substrate next to the metal (see line cuts in FIG \ref{fig: Transverse}(a)). Surprisingly, for the transverse voltage, the response changes sign when the laser crosses either the current-carrying or voltage probes. The sign and magnitude of $\mathrm{\Delta V_{xy}}$ showed no dependence on the incident laser polarization, strongly suggesting a thermal rather than non-linear origin.  

To investigate the origin of this response, we first repeated the measurements with a different laser wavelength. Since platinum exhibits no electronic or phononic excitations in the MIR, a nonlinear response would be of comparable size for 4.8 $\mathrm{\mu}$m and 10.6 $\mathrm{\mu}$m lasers. However, FIG \ref{fig: Transverse}(b) reveals an order of magnitude larger $\mathrm{\Delta \rho_{xy}}$ response from the longer wavelength light. For direct experimental comparison, we calculate the resistivity change by accounting for spot size variations, incident laser power, and current density: $\mathrm{\Delta\rho_{xy}\ =\ \frac{\Delta V_{xy}}{Current\ Density\ \times\ Fluence}}$. The 10.6 $\mathrm{\mu m}$ wavelength is close to the IR active phonon mode in the SiO$\mathrm{_2}$.\cite{kitamura2007optical} Thus, the large increase in $\mathrm{\Delta \rho_{xy}}$ is consistent with the enhanced absorbance in SiO$\mathrm{_2}$ producing a greater temperature rise in the platinum. These results suggest that the observed $\mathrm{\Delta \rho_{xy}}$ originates from a photothermal effect. 

Since $\mathrm{\Delta \rho_{xy}}$ changes sign with position and results from heating, it is tempting to assign it to a photo-Seebeck effect. However, since we are measuring across the vertical direction, a Seebeck response would only produce a sign change upon the laser crossing the horizontal probes. Changes in the x-position of the laser should only affect the magnitude of the response, not its sign. Thus, the sign change in both x and y suggests that it results from a current-induced second-order response to the thermal gradient (e.g., $\mathrm{\Delta V_{xy}\propto(\nabla_{x}T)(\nabla_{y}T)}$. If true, we would expect a nonlinear dependence on the laser power.

To check for nonlinearity, we fix the laser at the peak position of the response and independently sweep the applied current and laser power. The results, plotted in FIG \ref{fig: Transverse}(c), show that the transverse photovoltage, similar to the longitudinal response, depends linearly both on the current and laser power, ruling out any nonlinear origin. The linear fit to the power dependence, shown by a solid line in FIG \ref{fig: Transverse}(c),  yields a responsivity of 1.18 $\times~\mathrm{10^{-4}\ V\ W^{-1}}$ at a bias of 5 mA, corresponding to a power-normalized photoresistivity of $\mathrm{\rho^{(3)}_{xy}\ =\ 0.0686\ n\Omega\ mm^4\ W^{-1}}$.

To further check for subtle nonlinear thermal effects, we measured the dependence of the response on the chopper harmonics. Specifically, the laser should modulate the heat at the chopper frequency ($\mathrm{T[\omega]}$) so that a nonlinear thermal response would be observed at $\mathrm{2\omega}$ (i.e., $\mathrm{\Delta V_{xy}[2\omega]\propto(\nabla_{x}T[\omega])(\nabla_{y}T[\omega])}$). As shown in FIG \ref{fig: Transverse}(d), we observe an order of magnitude larger response at $\mathrm{\omega}$ compared to the higher harmonics, consistent with a linear response. Furthermore, we find that the higher harmonics are an experimental artifact. Indeed, the second and third harmonic responses are well modeled by considering the modulation of the Gaussian beam profile when it passes through a chopper with a finite thickness and angle of incidence (see the supplementary material).

To confirm that the photothermal origin of $\mathrm{\Delta V_{xy}}$ is also bolometric, we again measured its dependence on the modulation frequency. As shown in FIG \ref{fig: Transverse}(e), we observe a nearly frequency-independent response until 500 Hz, at which point the signal continuously decays. While we find the data is well fit to the standard bolometric response ($\mathrm{(1 + (2\pi f\tau)^2)^{-1/2}}$) with a thermal relaxation time of $\mathrm{\tau = 23.90~\mu}$s,\cite{richards1994bolometers} we are not able to chop fast enough to observe a substantial enough decrease in signal ($1/\sqrt{2}$) to truly determine the relaxation time. As such, we conclude the true relaxation time is likely much faster. Nonetheless, this response is orders of magnitude faster than we observed in $\mathrm{\Delta V_{xx}}$ response. 

To confirm that the much faster transverse response is indeed from the thermal gradient and bolometric type effect, we again performed additional COMSOL simulations. As shown in FIG \ref{fig: Transverse}(f), upon terminating the heat source, the simulated $\mathrm{\Delta V_{xy}^{sim}}$ exhibits an exponential decay with $\mathrm{\tau = 0.44~\mu}$s. Observing such time scales would require modulation frequencies up to a megahertz, well beyond our current setup. Nonetheless, the transverse response's orders of magnitude shorter relaxation time than that of the longitudinal is consistent with a distinct mechanism. Specifically, the transverse signal depends on local temperature gradients that equilibrate rapidly, while the longitudinal response is governed by the overall device temperature that relaxes more slowly through the substrate.

\begin{figure}[h]
    \centering
    \includegraphics[width=.9\linewidth]{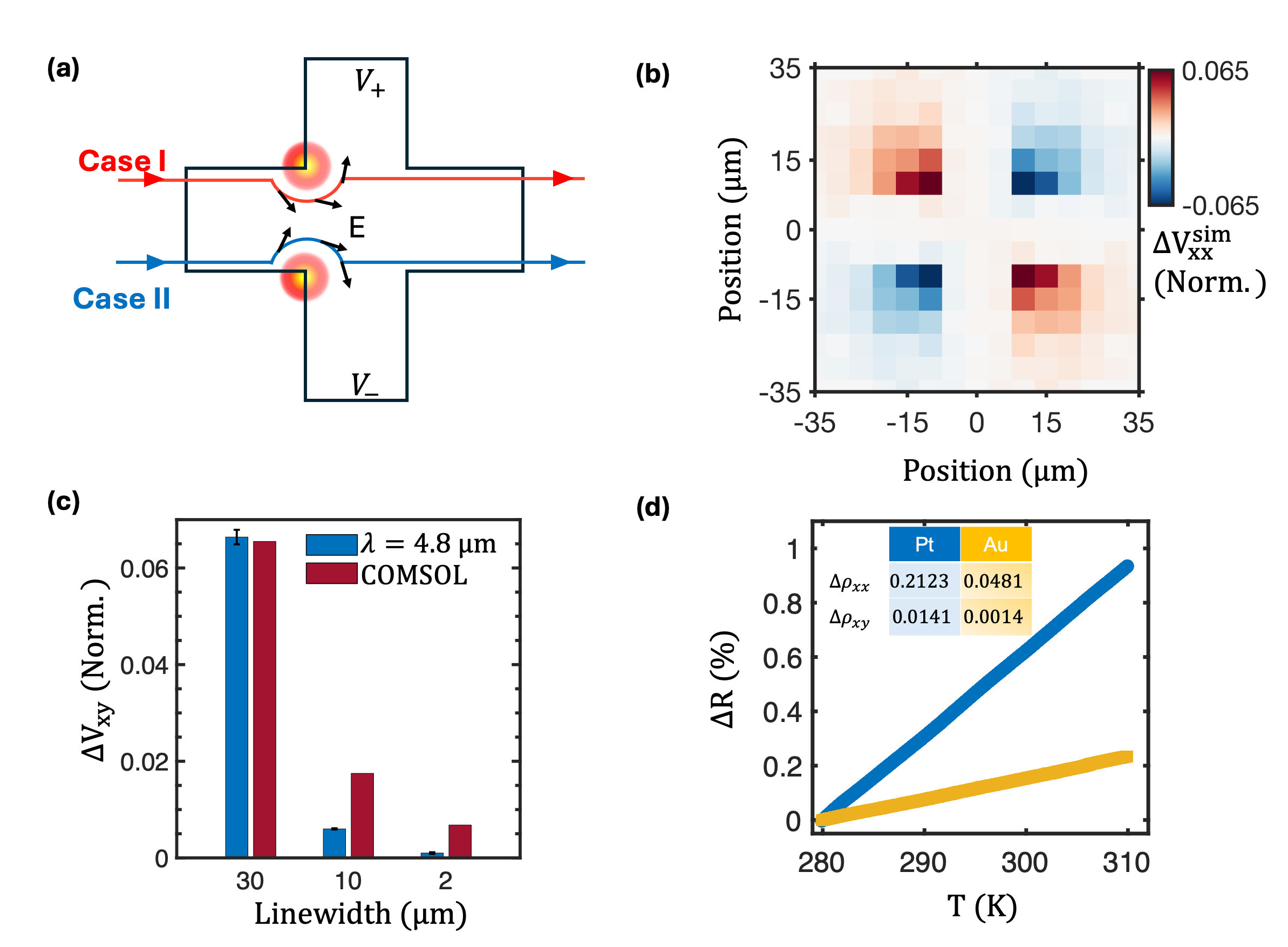}
    \caption{(a) 
Schematic: a MIR-induced local hot spot near an edge increases resistivity and deflects current, producing $\mathrm{\Delta V_{xy}}$ whose sign flips when the hot region crosses a probe axis    
    (b) COMSOL simulation map of the normalized transverse photovoltage. (c) Histograms of the peak transverse response for different widths of the 30 $\mathrm{\mu}$m platinum device. Error bars represent the standard deviation of peak values measured across current hotspots. (d) Percent change in resistivity from room temperature to 310 K for platinum and gold devices. Inset: Table of photoresistivity of transverse and longitudinal in units of $\mathrm{n\Omega.~mm^4.~W^{-1}}$. The larger resistivity changes in platinum with temperature produce a greater photothermal response.}
    \label{fig: 4}
\end{figure}

Having established that both effects are bolometric in nature, we now turn to understand the origin of their discrepant behaviors. Contrary to $\mathrm{\Delta V_{xx}}$, $\mathrm{\Delta V_{xy}}$ has a much faster response, is localized near the sample edges, and changes sign upon crossing the device. The laser primarily heats the SiO$_2$/Si substrate via photon–phonon absorption, with subsequent thermal diffusion into the metal film. This localized heating induces resistance changes according to $\mathrm{\Delta R/R = \alpha_T \Delta T}$, where the temperature coefficient of resistance $\mathrm{\alpha_T = (1/R)(dR/dT)}$, directly generates the observed photovoltage response.
This region of higher resistance causes the otherwise uniform current to deflect, as illustrated in FIG \ref{fig: 4}(a). First, we consider Case I, where the laser is in the top-left corner. Here, the photovoltage is positive as the probe measures the region where the current (red line) is deflected upward. When the laser is moved to the bottom-left corner (Case II), the voltage is negative, as it now probes the region where the current (blue) deflects downward. Similarly, placing the laser at the top (bottom) right results in probing the region where the current is deflected down (up), leading to negative (positive) voltages.

To confirm the current deflection as the origin of the transverse voltage, we performed additional COMSOL simulations and experiments. As shown in FIG \ref{fig: 4}(b), our simulations find a similar switch in the sign of the transverse voltage upon moving the heat across the x or y probes that depends on the induced temperature gradient. We note that the simulations do not directly include light absorption and thus only capture the effect of heating and the resulting local change in resistance. We further checked experimentally that the transverse voltage depends on the temperature gradient. Specifically, by changing the device width while keeping the laser spot size fixed, we alter the size of the thermal gradient sampled and thus the magnitude of the current deflection. 

To this end, we tested the devices with 3 different widths: 2 $\mathrm{\mu}$m, 10 $\mathrm{\mu}$m, and 30 $\mathrm{\mu}$m. All other dimensions, including the cross length and thickness, were kept constant. The measured and simulated peak responses for each width are shown in FIG \ref{fig: 4}(c). We observe a clear trend of decreasing transverse response amplitude upon reducing the device width, consistent with our COMSOL simulations. The underestimation of the change in $\mathrm{\Delta V_{xy}}$ with width likely results from unmodeled disorder and grain-boundary scattering reducing the effective $\mathrm{\alpha_T}$. Within experimental uncertainty $\mathrm{\rho^{(3)}_{xx}\ \sim \ 0.22\ n\Omega .\ mm^4\ .W^{-1}}$, at maximum response, is the same for all three devices. This width independent behavior of $\mathrm{\Delta V_{xx}}$ is also confirmed by simulations.

For future measurements on centrosymmetric materials utilizing the same measurement scheme and device design, in an attempt to reduce the effect, we turned to Au devices. Au has a bulk thermal conductivity $\kappa$ of 318 W m$^{-1}$ K$^{-1}$, which is four times that of Pt (72 W m$^{-1}$ K$^{-1}$).\cite{ho1972thermal} The temperature-dependent resistance change near room temperature, as shown in FIG \ref{fig: 4}(d), is significantly higher for Pt thin films. To quantify this, we extracted $\mathrm{\alpha_T}$  from the slope of the linear fits, yielding $\mathrm{3.12\ \times 10^{-4}~ K^{-1}}$ for Pt and $\mathrm{7.78\ \times 10^{-5}\ K^{-1}}$ for Au. Consequently, we expect a smaller photothermal response in Au, since the effect originates from temperature-induced resistivity changes. Indeed, as shown in the inset table of FIG \ref{fig: 4}(d), the ratio of longitudinal photoresistivities between Pt and Au nearly matches the ratio of their $\mathrm{\alpha_T}$ values. The transverse photoresistivity is even more strongly suppressed in Au, likely due to its higher thermal conductivity, to which the transverse response is particularly sensitive.

%%%%%%%%%%%%%%%%%%%%%%% Conclusion SECTIONS %%%%%%%%%%%%%%%%%%%%%%%%%%%%%%%%%%%%%%
Our results reinforce the importance of careful device design and measurement techniques that can be exploited to minimize extrinsic thermal effects. As the community moves toward increasingly complex measurements, it is crucial to be aware of extrinsic photo-induced electronic signals and how they contribute to photocurrent measurements. Thermal effects, in particular, are difficult to avoid, especially when using IR light sources. We have demonstrated that even when devices are designed to eliminate Seebeck contributions, thermal current deflection presents a significant challenge for field-induced nonlinear measurements. This effect is amplified in materials with large temperature coefficients of resistance, as is the case for many topological semimetals.\cite{kumar2017extremely,yang2021evidence,sankar2015large}

Based on our findings, we propose several strategies to minimize extrinsic optoelectronic responses. First, symmetric cross-shaped geometries with extended contact regions effectively eliminate photo-Seebeck contributions by ensuring thermal gradients remain far from the contact interfaces. Second, material selection proves critical: materials exhibiting weaker $\mathrm{\alpha_T}$  at operational temperatures will show reduced bolometric effects, enabling clearer observation of intrinsic responses. The metal contacts with lower temperature coefficients of resistance and higher thermal conductivity, such as gold ($\mathrm{\alpha_T}$ = $7.78\ \times 10^{-5}\ K^{-1}$, $\kappa$ = 318 W m$^{-1}$ K$^{-1}$) compared to platinum ($\mathrm{\alpha_T}$ = $3.12\ \times 10^{-4}\ K^{-1}$, $\kappa$ = 72 W m$^{-1}$ K$^{-1}$), significantly reduce photothermal signals. Third, device dimensions should be optimized based on the specific measurement. Our results demonstrate that narrower devices (2 $\mu$m width) minimize transverse thermal artifacts by limiting current deflection, while the longitudinal response remains insensitive to width variations. Fourth, operating at chopping frequencies close to or above the thermal roll-off can suppress both longitudinal and transverse thermal contributions. Finally, employing substrates that are fully transparent at the wavelengths of interest eliminates substrate-mediated thermal contributions, isolating the intrinsic material response. For measurements requiring IR excitation where substrate absorption is unavoidable, thermally conductive substrates with minimal absorption (e.g., sapphire) should replace SiO$_2$/Si to enhance heat dissipation. Indeed, we found fabricating devices on sapphire resulted in suppressing the transverse response.

Despite these challenges, the current deflection mechanism observed here could enable novel device applications. By incorporating metamaterials with engineered polarization responses into the thermally active regions, one could create polarization-sensitive detectors. Furthermore, the device geometries and measurement protocols developed here can be adapted to explore Nernst and nonlinear thermoelectric effects in topological systems, where these design principles prove particularly valuable for isolating intrinsic quantum geometric contributions from extrinsic thermal backgrounds.
\section*{SUPPLEMENTARY MATERIAL}
See the supplementary material for device fabrication, photovoltage measurement, and COMSOL simulation details. The analysis of chopper harmonics dependence of the photovoltage response is also included.

\section*{ACKNOWLEDGMENTS}
The photocurrent experiments, analysis, device design, fabrication, and experimental optimization were supported by the US Department of Energy (DOE), Office of Science, Office of Basic Energy Sciences under award number DE-SC0018675 (VMP, PS, GN, MG, CG, KSB). DR is grateful for the Undergraduate Research Fellowship at Boston College. We are grateful for fruitful discussions with Hou-Tong Chen. 

%%%%%%%%%%%%%%%%%%%%%%% Author Contribution %%%%%%%%%%%%%%%%%%%%%%%%%%%%%%%%%%%%%%

\section*{AUTHOR DECLARATIONS}
\subsection*{Conflict of Interest}
The authors have no conflicts to disclose.
\subsection*{Author Contributions}
VMP, PS, and CG performed, optimized, and analyzed the photovoltage experiments. PS performed the temperature-dependent resistance measurement. PS fabricated the devices with help from MG and GN. DR performed the COMSOL simulations under the supervision of KK. YR, KSB, and KK proposed the phenomenological explanation. VMP and PS wrote the manuscript with the help of KSB. KSB conceived and supervised the project. All authors contributed to the discussion of the manuscript.

\subsection*{DATA AVAILABILITY}
The data that support the findings of this study are available from
the corresponding author upon reasonable request. 
\bibliography{Bibliography}
\end{document}